\documentclass{sig-alternate}

\usepackage{latexsym}
\usepackage{amsmath}
\usepackage{amssymb}
\usepackage{epsfig}

\newtheorem{theorem}{Theorem}[section]
\newtheorem{corollary}[theorem]{Corollary}
\newtheorem{lemma}[theorem]{Lemma}

\newtheorem{definition}[theorem]{Definition}

\newtheorem{example}[theorem]{Example}
\newtheorem{remark}[theorem]{Remark}

\newcommand{\comment}[1]{}
\newcommand{\removeLinSCPM}[1]{}
\newcommand{\newExpSCPM}[1]{#1}
\newcommand{\oldExpSCPM}[1]{}
\newcommand{\V}[1]{\vec{#1}}
\newcommand{\score}{{\cal{S}}}

\def\Loss{{\mathcal L}}
\def\minimize{\mathop{\rm minimize}}
\def\maximize{\mathop{\rm maximize}}
\def\subto{\rm{ s.t.}}
\newcommand{\quoteIt}[1]{``#1''}

\def\Expect{{\mathbb E}}

\newcommand{\argmin}{\mathop{\rm argmin}}

\begin{document}

\title{A Unified Framework for Dynamic Pari-Mutuel Information Market Design}
\subtitle{[Extended Abstract]}
%
%
%
%
%

\numberofauthors{5} 
%
\author{
%
%
\alignauthor Shipra Agrawal \email{shipra@stanford.edu}
\alignauthor Erick Delage \email{edelage@stanford.edu}
\alignauthor Mark Peters \email{mark\_peters@stanford.edu}
\and
\alignauthor Zizhuo Wang \email{zzwang@stanford.edu}
\alignauthor Yinyu Ye \email{yyye@stanford.edu}
}
\additionalauthors{Additional authors: John Smith (The Th{\o}rv{\"a}ld Group,
email: {\texttt{jsmith@affiliation.org}}) and Julius P.~Kumquat
(The Kumquat Consortium, email: {\texttt{jpkumquat@consortium.net}}).}
\date{30 July 1999}

\maketitle


\begin{abstract}
Recently, several new pari-mutuel mechanisms have been introduced to
organize markets for contingent claims.
Hanson \cite{Hanson} introduced a market maker derived from the logarithmic scoring rule, and later
Chen \& Pennock \cite{yiling} developed a cost function formulation for the market maker. On the other hand,
the SCPM model of Peters et al. \cite{peters1} is based on ideas from a call
auction setting using a convex optimization model.
In this work, we develop a unified framework that bridges these
seemingly unrelated models for centrally organizing contingent claim
markets.  The framework, developed as a generalization of the SCPM,
will support many desirable properties such as proper scoring,
truthful bidding (in a myopic sense), efficient computation, and
guarantees on worst case loss.
In fact, our unified framework will allow us to express
various proper scoring rules, existing or new,
from classical utility functions in a convex optimization problem representing
the market organizer. Additionally, we utilize concepts from duality to show that the
market model
is equivalent to a risk minimization problem where a convex risk measure is employed.
This will allow us to more clearly
understand the differences in the risk attitudes adopted by various mechanisms, and particularly deepen
our intuition about popular mechanisms like Hanson's market-maker.
In aggregate, we believe this work advances our understanding of the
objectives that the market organizer is optimizing in popular
pari-mutuel mechanisms by recasting them into one unified framework.
\end{abstract}




\section{Introduction}
Contingent claim markets are organized for a variety of purposes.
Prediction markets are created to aggregate information about a
particular event.  Financial markets involving contingent claims
allow traders to hedge their exposure to certain event outcomes.
Betting markets are designed for entertainment purposes.  The
participants in these markets trade securities which will pay a
fixed amount if a certain event occurs.  Some examples of these
events would be the winner of the World Series, the value of the
latest consumer price index or the release date of Windows Vista.
Prediction markets have grown in popularity as research into the
accuracy of their predictions has shown that they effectively
aggregate information from the trading population.  One of the
longest-running prediction markets is the Iowa Electronic Market
which allows real money betting on various elections. Studies by
Berg and her coauthors (see \cite{BR06}, \cite{BNFR08} and
\cite{BNR08}) have shown that the information generated by these
markets often serves as a better prediction of actual outcomes than
polling data. Google has run internal prediction markets over a
variety of events and Cowgill et al. \cite{CWZ08} have shown that
their predictions also perform quite well.\\
\newline
Despite the potential value created by these markets, there can be
some difficulties with their introduction and development.  First,
many nascent markets suffer from liquidity problems. Occasionally
these problems stem from the choice of mechanism used to operate the
market. Organizing markets as a continuous double auction (like the
NASDAQ stock market) is a popular option and usually performs well.
However, in thin markets, Bossaerts et al. \cite{BFL02} have
demonstrated that some problems surface which inhibit the growth of
liquidity.  To overcome this situation, the market organizer could
introduce an automated market-maker which will centrally interact
with the traders.  This mechanism will have some rules for pricing
shares. The market organizer must determine these rules with one key
question being his own tolerance for risk.  Recently, there has been
a surge in research of these automated market-makers.\\
\newline
New market-making mechanisms based on pari-mutuel principles have
recently been developed by Hanson \cite{Hanson}, Pennock et al.
\cite{PCD06} and Peters et al. \cite{peters1}.  These market-makers
allow contingent claims in nascent market to be immediately priced
according to rules of the mechanism.  The mechanisms are pari-mutuel
in the sense that 
the winners are generally paid out by the stakes of the losers.  The
claims being traded are commitments to pay out a fixed amount if a
particular event occurs in the future. The mechanism developed by
Hanson has been shown to perform well in simulated markets
\cite{peters} and has been adopted
by many online prediction markets.\\
\newline
However, the origins of these new mechanisms differ.  Peters et al.
\cite{peters1} developed their mechanism by creating a sequential
version of a call auction problem which is solved by convex
optimization. Their Sequential Convex Pari-mutuel Mechanism (SCPM)
uses an optimization problem to determine when to accept orders and
how to price accepted orders. On the other hand, Hanson's mechanism
is derived from scoring rules. Scoring rules are functions often
used to compare distributions.  In particular, Hanson uses the
logarithmic scoring rule to determine how much to charge a trader
for a new order.
His mechanism is called the Logarithmic Market Scoring Rule (LMSR).  Using
a similar approach as Hanson, it is possible to create market-makers
for other scoring rules.
In contrast to the SCPM, the MSR model doesn't directly provide an
optimization problem from the market organizer's standpoint. \\
\newline
Recently, there has been some interest in comparing and unifying
these mechanisms for prediction markets. Chen and Pennock
\cite{yiling} give an equivalent cost function based formulation for
the MSR market makers, and relate them to utility-based market
makers. They show that a certain class (hyperbolic absolute risk
aversion) of utility-based market is equivalent to a market scoring
rule market maker. Peters et al. \cite{peters} empirically compare
the performances of various market mechanisms. In this work, we
provide strong theoretical foundation for unifying
existing market makers like the SCPM, the MSR and cost function
based markets under a single convex optimization framework. Our
model not only aids in comparing various mechanisms, but also
provides intuitive understanding of the behavior of the market
organizer in these seemingly different mechanisms. Specifically, our
main contributions are as follows:
\begin{itemize}
\item {\em The unifying framework}: we propose a generalized version of the SCPM as a unified convex optimization framework for market makers.
The new model enjoys many desirable properties like a truthful pricing scheme, efficient cost function formulation, scoring properness, and
guarantees on worst case loss bounds.
\item {\em Equivalence to the MSR market makers:} we establish our claim that the new SCPM framework unifies existing market makers by showing that
a) any market maker based on a proper scoring rule (MSR) can be formulated as a special case of this SCPM model; and conversely, b)
an objective function in our framework
implicitly corresponds to a (strictly) proper scoring rule under certain easily verifiable conditions.
\item {\em Intuitive interpretation of the market mechanisms:}  We show that the market maker's model in the new framework is equivalent to
convex risk minimization for the market maker. Thus, when using this
model to accept incoming orders the market maker is actually taking
rational decisions with respect to a risk attitude as defined by the
choice of utilities of the framework. Further, we show that for some
popular mechanisms like the LMSR, the implicit risk function turns
out to characterize precisely how much the market maker is prepared
to invest in order to learn a distribution $\V{p}$ which is very
different from his prior belief.
This explains the design choices of these mechanisms.
\item {\em New mechanisms:} We take a step forward and demonstrate with examples that various insights provided by the unified framework
can be used to guide the design of new mechanisms that have desirable properties with respect to various criteria discussed:
truthfulness, worst case loss, computational efficiency, risk attitude, properness of the corresponding scoring rule, etc.
\end{itemize}

The rest of the paper is organized as follows. To begin, Section
\ref{background} will provide some background on the mechanisms
which we will be studying.  In Section \ref{sec:unifying}, we
propose our new framework based on a generalization of the SCPM, and
demonstrate its properties of truthful pricing scheme, cost function
based formulation, and guarantees on worst case loss bounds. In
Section \ref{sec:unified}, we show that the SCPM framework is
equivalent to the MSR market makers. Section \ref{risk} will further
explain how the market organizer's decision problem in our unified
framework is actually
equivalent to a convex risk minimization problem. 
We conclude the paper with a discussion in Section \ref{discussion}
where we compare the existing mechanisms, and provide examples
illustrating
simple guidelines for designing new ones.

\section{Background} \label{background}
In this section, we will provide some background on the key
mechanisms for prediction markets discussed in this work -- the
market scoring rule mechanisms (MSR), cost-function formulation of
markets, and the Sequential Convex Pari-mutuel mechanism (SCPM).

Let $\omega$ represent a discrete or discretized random event to be
predicted, with $N$ mutually exclusive and exhaustive outcomes. We
consider a contingent claims market where claims are of the form
``Pays \$1 if the outcome state is $i$".
A new trader arrives and submits an order which essentially specifies the claims over each outcome state that the trader desires to buy.
The market maker then decides what price to charge for the new order. Various mechanisms treat a new order in the following seemingly
different manners:

\subsection{Market Scoring Rules}
\label{background-MSR}
Let $\V{r} = (r_1, r_2, ..., r_N)$ represents a probability estimate
for the random event $\omega$. A scoring rule is a sequence of
scoring functions, $S = \score_1(\V{r}), \score_2(\V{r}),
...,\score_N(\V{r})$, such that a score $\score_i(\V{r})$ is
assigned to $\V{r}$ if outcome $i$ of the random variable $\omega$
is realized. A proper scoring rule \cite{winkler1969} is a scoring
rule that motivates truthful reporting. Based on proper scoring
rules, Hanson\cite{Hanson} developed a Market Scoring Rule(MSR)
mechanism. In the MSR market, the market maker with a proper scoring
rule $\score$ begins by setting an initial probability estimate,
$\V{r}_0$. Every trader can change the current probability estimate
to a new estimate of his choice as long as he agrees to pay the
market maker the scoring rule payment associated with the current
probability estimate and receive the scoring rule payment associated
with the new estimate.

Some examples of market scoring rules are the logarithmic market
scoring rule (LMSR)\cite{Hanson}: 
$$\score_i(\V{r}) = b\log(r_i) \ \ \ \ \  (b>0)$$
and the quadratic scoring rule:
$$\score_i(\V{r}) = 2b r_i -b \sum_j r_j^2 \ \ \ \ \ (b>0)$$
Hanson's MSR has many favorable characteristics. It is designed as a
pari-mutuel mechanism which bounds the risk of the market organizer.
It functions as an automated market maker in the sense that it is
always able to calculate prices for new orders. Crucially, the LMSR
is also known to elicit {\it truthful} bids from the
market traders.

\subsection{Cost function of Market-makers}
\label{sec:background-cost-function}
Recently, Chen \& Pennock \cite{yiling} proposed a cost function based implementation
of market makers. 
Let the vector $\V{q} \in{\mathbb
R}^N$ represents the number of claims on each state currently held by the traders.
In the cost function formulation, the total cost of all
the orders $\V{q}$ 
is calculated via some cost function $C(\V{q})$.  A trader submits
an order characterized by the vector $\V{a} \in{\mathbb R}^N$ where
$a_i$ reflects the number of claims over state $i$ that the trader
desires. The market organizer will charge the new trader
$C(\V{q}+\V{a}) - C(\V{q})$ for his order.
At any time in the market, the going price of a claim for state $i$, $p_i(\V{q})$, equals $\partial C/\partial q_i$. The price is
the cost per share for purchasing an infinitesimal quantity of security $i$.\\
\newline
Chen \& Pennock \cite{yiling} show that any scoring rule has an equivalent cost function formulation.
For example, below are the specific cost and pricing functions for LMSR:
\begin{equation*}
\begin{array}{lll}  
C(\V{q}) = b \log{ \left( \sum_j e^{q_j/b} \right)} & \mbox{ and } & p_i(\V{q}) = \frac{e^{q_i/b}}{\sum_j e^{q_j/b}}
\end{array}
\end{equation*}
For general market scoring rules, they proposed three equations that the cost
function $C$ should satisfy so that the cost function based market maker is equivalent to the market based on a given scoring rule $\score$:
\begin{equation}
\begin{array}{l}\label{score-cost}
\score_i(\V{p})=q_i-C(\V{q})+K\mbox{       }\forall i\\
\sum_ip_i=1 \\
p_i=\frac{\partial C}{\partial q_i}\mbox{       }\forall i\\
\end{array}
\end{equation}
We will use this formulation later to prove equivalence of MSR and SCPM mechanisms.

\subsection{Sequential Convex Pari-Mutuel Mechanism}
The SCPM was designed to require traders to submit orders which
include three elements: a limit price ($\pi$), a limit quantity
($l$) and a vector ($\V{a}$) that represents which states the order
should contain. The components of the vector $\V{a}$ will contain
either a $1$ (if a claim over the specified state is desired) or a
$0$ (if it is not desired). The limit price refers to the maximum
amount that the trader wishes to pay for one share.  The limit
quantity represents the maximum number of shares that the trader is
willing to buy. The market maker decides the actual number of shares
$x$ to be granted to a new order, and the price to be charged for
the order. The market maker solves the following optimization
problem for making this decision:
\begin{equation}\label{scpm_marks}
\begin{array}{c@{\quad}l@{\qquad}l}
\mbox{maximize}_{x,z,\V{s}} & \pi x - z + \sum_i {\theta}_i \log({s}_i) \\
\noalign{\smallskip}
\mbox{subject to} & \V{a} x + \V{s} + \V{q} = z\V{e} & \\
\noalign{\smallskip}
& 0\le x \le l,\\
\end{array}
\end{equation}
where parameters $\V{q}$ stands for the numbers of shares currently held
by the traders prior to the new order $(\pi,l,\V{a})$ arrives, and $\V{e}$ represents the vector of all 1s.
Each time a new order arrives, the optimization problem (\ref{scpm_marks}) is solved and the state prices are defined to be
the optimal dual variables corresponding to the first set of constraints denoted as $\V{p}$. The trader is then charged according to the
inner product of the final price and the order filled $(\V{p}^T\V{a})$. \\

This optimization problem, without the utility function $\sum_i \theta_i \log s_i$, has the following interpretation for the market maker: besides $x$, decision variable $z$ represents the maximum number of accumulated shares, including $\V{s}$ representing the contingent numbers of surplus shares would be kept by the market maker,
over all states; and $(\pi x-z)$ in the objective represents the profit
that would be made from the new order. Thus, market organizer aims at maximizing his worst case profit.
As we establish later in this paper, adding the utility function $\sum_i {\theta}_i \log({s}_i)$ enhances the risk taking ability of the market maker.

\section{The unifying framework}
\label{sec:unifying}
In this paper, we illustrate that a generalized formulation of SCPM provides a unifying framework for
pari-mutuel market mechanisms. We propose the following convex
optimization model with
a {\em{concave}} continuous utility function $u(\V{s})$
\footnote{Indeed, there is another technical
condition on $u(\cdot)$ that is required for this model to be
feasible and bounded, that is, $\forall \V{q} \ge 0$, $\exists t$,
$\nabla u(t\V{e}-\V{q})^T\V{e}=1$, where $\nabla u(\cdot)$ denotes the (sub-)gradient function.}:
\begin{equation}
\begin{array}{ll}\label{scpm1}
\mbox{maximize}_{x,z,\V{s}}& \pi x-z+u(\V{s})\\
\mbox{subject to}& \V{a} x + \V{s} + \V{q}=z\V{e}\\
&0\leq x\leq l\\
\end{array}
\end{equation}
\newline
Note that utility function $u(\V{s})=\sum_i \theta_i \log s_i$ used in the original SCPM model of Peters et al. \cite{peters} is a special case
of (\ref{scpm1}).
From here on, ``SCPM'' refers to the above generalized SCPM model. When required, we disambiguate by referring to the original model with
$u(\V{s})=\sum_i \theta_i \log s_i$  as ``Log-SCPM''. The optimization model (\ref{scpm1}) has exactly the same meaning for the market maker as the
Log-SCPM, and inherits many desirable properties like intuitive interpretation, convex formulation, global optimality, Lagrange duality,
polynomial computational complexity, etc.,
as in the original Log-SCPM model.
Next, we demonstrate some new desirable properties of the new SCPM framework including truthfulness of the pricing scheme, efficient cost function
based scoring rules, and easily computable guarantees on worst case loss.
\subsection{Truthful pricing scheme}
\label{sec:truthfulness}
The original SCPM model does not provide incentives for the traders to bid truthfully \cite{peters}.
On the other hand, market scoring rules such as LMSR ensure truthful bidding. We show that this difference in
incentives is attributed to a difference between the implementation of SCPM and MSR pricing scheme.
In the SCPM model, the market organizer will
typically charge the trader for an accepted number of shares based
on the final price calculated by the mechanism. However, in the
market scoring rules such as LMSR, the trader is actually charged by a cost function which is
equivalent to the integral of the pricing function over the number
of shares accepted. Thus, as the price increases while the order is
filled, the trader is charged the {\em instantaneous price} for each
infinitesimally small portion of his order that is filled.

We show that our general SCPM framework equally admits truthfulness
if the market maker charge traders according to the integral of the
price function over infinitesimally small accepted shares. More
specifically, the new charging scheme solves the optimization
problem for infinitesimally small orders, thus computing incremental
prices. Define $\V{p}(\epsilon)$ to be the dual prices corresponding
to the following optimization problem (with $l=\epsilon$):
\begin{equation}
\begin{array}{ll}\label{scpm2}
\mbox{maximize}_{x,z,\V{s}}& \pi x-z+u(\V{s})\\
\mbox{subject to}& \V{a} x + \V{s} + \V{q} = z\V{e}\\
&0\leq x \leq \epsilon\\
\end{array}
\end{equation}
\newline
The SCPM model (\ref{scpm1}) can be equivalently viewed
as gradually increasing $\epsilon$ until the price become more than the bid offered, or
the trader reaches his limit $l$.
Let $\bar{x}$ be the optimal solution
to (\ref{scpm1}), i.e. the largest $\epsilon$ such that the last
constraint in (\ref{scpm2}) is tight at optimality.
Now, instead of charging the trader the final price
$\V{p}(\bar{x})^T\V{a}$ as in the conventional SCPM, we charge an
accepted order by the following formula:
\begin{equation}
\begin{array}{l}\label{charge}
(\int_{0}^{\bar{x}}\V{p}(\epsilon)d\epsilon)^T \V{a}\\
\end{array}
\end{equation}

Below, we establish the integrability of the pricing function and some important properties of the new pricing mechanism:
\begin{lemma}\label{price-properties}
\begin{enumerate}
\item The price vector sums up to 1 for every $\epsilon$;
\item The price is non-negative if the utility function $u(\cdot)$ is non-decreasing;
\item The price is consistent, i.e., the market organizer will
    accept another infinitesimal order if $\pi$ is greater than
    the instantaneous price of the order and vice versa;
\item The instantaneous price $\V{p}(\epsilon)^T\V{a}$ is non-decreasing in $\epsilon$;
\item The price function is integrable.
\end{enumerate}
\end{lemma}
\begin{proof}
The proof of this theorem is provided in Appendix \ref{app:price-properties}.
\end{proof}
The above properties of the pricing scheme lead to the following strong result about the truthfulness of this scheme:
\begin{theorem}
Irrespective of the choice of utility function $u(\cdot)$, the optimal bidding strategy in the SCPM is myopically
truthful when the traders are charged according to pricing scheme (\ref{charge}).
\end{theorem}
\begin{proof}
Assume that the trader has a valuation of $\gamma$ for his desired
order $\V{a}$ with a quantity limit of $l$.  The trader's profit can
be expressed as: $R(x) = \gamma x - (\int_{0}^{x}
\V{p}(\epsilon)d{\epsilon})^T\V{a}$. The trader seeks to maximize $R$ by
choosing a proper $\pi$.
  From the previous lemma, we know that $\V{p}(\epsilon)^T\V{a}$ is
  non-decreasing in $\epsilon$. Thus, it is easy to see that an
optimal strategy when $\V{p}(0)^T\V{a} \geq \gamma$ is to not
place an order.  Thus, we will assume that $\V{p}(0)^T \V{a} < \gamma$.\\
Also, since $p(x)$ is non-decreasing in $x$, $R(x)$ is a concave function in $x$. Therefore, the optimal $x \le l$ (that maximizes trader's profit $R(x)$) is given by the following optimality conditions
\begin{equation*}
\begin{array}{l}
\left(x - l \right) \left( \gamma - \V{p}(x)^T\V{a} \right) = 0 \\
 \gamma \geq \V{p}(x)^T \V{a}
\end{array}
\end{equation*}

On the other hand, the market organizer will accept a bid if and only
if $\V{p}(\epsilon)^T\V{a}$ is less than $\pi$. Hence, setting
$\pi=\gamma$ gives the optimal
$x$, and thus optimal profit for the trader. 
This is equivalent to bidding truthfully
since the trader will essentially bid such that the instantaneous
price is driven either to their valuation or as close to their
valuation as possible before hitting the limit quantity constraint.
\end{proof}

Therefore, we have shown that the truthfulness of this mechanism does not depend on the particular choice of utility function
but rather on the implementation of the charging method. By revising the charging method to an `incremental' one, we can create a
truthful implementation for the general SCPM.  As we show in the next section, 
the integral in the expression (\ref{charge}) does not need to be explicitly calculated in order to compute it;
given $\bar{x}$, we can compute the total charge efficiently using a convex cost function formulation.

\subsection{Cost function of the market maker}
\label{sec:cost-function-equiv}
Section \ref{sec:background-cost-function} discussed a cost function based implementation for market makers, introduced by Chen and Pennock \cite{yiling}.
In this section, we derive a convex cost function for the SCPM, which will reduce the problem of computing the integral (\ref{charge}) to a
simple convex optimization problem.
\begin{theorem}
\label{th:SCPM-cost-function}
Let $\V{q}$ be the number of shares on each state held by the traders in the SCPM market, and
a new order $(\pi,l,\V{a})$ is accepted up to level $\bar{x}$ and is
charged according to pricing scheme (\ref{charge}). Then, the charge
is
\[\left(\int_{0}^{\bar{x}}\V{p}(\epsilon)d\epsilon\right)^T \V{a}=C(\V{q}+\V{a}\bar{x})-C(\V{q}),\]
where $C(\V{q})$ is a convex cost function defined by
\begin{equation}\label{costofscpm}
C(\V{q}) = \min_t t - u(te-\V{q}),
\end{equation}
and has the property $\V{e}^T\nabla C(\V{q})=1,\ \forall \V{q}\ge 0$.
\end{theorem}

\begin{proof}
Note that $\V{p}(\epsilon)$ is the optimal dual solution associated with the optimization problem:
\begin{eqnarray*}
\maximize_{z,\V{s}} && \pi \epsilon - z + u(\V{s})\\
\subto && \V{a}\epsilon + \V{s} + \V{q} = z\V{e}.
\end{eqnarray*}
Let $V(\V{q},\epsilon)$ denote the optimal objective value of the above problem. Then,
\begin{eqnarray*}
V(\V{q},\epsilon) & =&
\begin{array}{rl}
\max_{z,\V{s}}& \pi \epsilon - z + u(\V{s})\\
\subto & \V{a}\epsilon + \V{q} + \V{s} = z\V{e}
\end{array} \\
&=& \pi \epsilon -\min_z \Big\{z-u(z\V{e}-\V{q}-\V{a}\epsilon)\Big\} \\
& = & \pi\epsilon - C(\V{q}+\V{a}\epsilon)\nonumber
\end{eqnarray*}
Next, using local sensitivity analysis results (e.g., discussed in \cite{hindi:tutorial}), the optimal dual variables
$\V{p}(\epsilon)$
$$
p(\epsilon)_i = -\frac{\partial V(\V{q},\epsilon)}{\partial q_i} = \frac{\partial C(\V{q}+\V{a}\epsilon)}{\partial q_i}
$$
Thus, we can conclude the statement presented in our theorem by performing the integration
\begin{eqnarray*}
\int_0^{\bar{x}} \V{p}(\epsilon)^T \V{a} d\epsilon & = & \int_0^{\bar{x}} \nabla_q C(\V{q}+\V{a}\epsilon)^T \V{a} d\epsilon \\
& = & \int_0^{\bar{x}} \frac{dC(\V{q}+\V{a}\epsilon)}{d\epsilon}d\epsilon \\
& = & C(\V{q}+\V{a}\bar{x})-C(\V{q})
\end{eqnarray*}
Finally, we can verify that $C(\V{q})$ is a convex function of $\V{q}$ since
it is the minimum over $t$ of a function that is jointly convex in
both $t$ and $\V{q}$.
\end{proof}
Note that not only is the cost function convex, but it can also be
simply computed for any given $\V{q}$ by solving a single variable
convex optimization problem. This is in contrast to the market
scoring rules, where computing the cost function is non-trivial and
requires solving a set of differential equations (refer equation
(\ref{score-cost}) in section
\ref{sec:background-cost-function}).


\subsection{Worst case loss for the market maker}
\label{sec:loss}
An interesting consequence of the cost function representing the SCPM
developed in the previous section, is that the worst case loss can be formulated as a convex optimization problem:
\begin{theorem}
\label{th:loss}
Assuming the market starts with $0$ shares initially, then the worst case loss for the market maker using the SCPM mechanism
is given by $B+C(0)$ where
\begin{equation*}
\begin{array}{lll}
B & = & \max_{i}\{ \max_{\V{s}} \  u(\V{s}) -s_i \}
\end{array}
\end{equation*}
and $C(\cdot)$ is the cost function defined by (\ref{costofscpm}).
\end{theorem}
\begin{proof}
Let the number of shares held by the traders at time $t$ is
$(\V{q})^t$.
By Theorem~\ref{th:SCPM-cost-function},
assuming we started with $0$ shares initially, the total money
collected at time $t$ is $C((\V{q})^t) - C(0)$.
%
On the other hand, if state scenario $i$ occurs, the market maker
needs to pay amount $(q_i)^t$,
Thus, we can find for each state scenario
$i$, the worst case loss by solving the optimization problem :
\begin{eqnarray*}
\max_{\V{q}\ge 0} q_i-(C(\V{q})-C(0)) & = & \max_{t, \V{q}\ge 0}\{(q_i -t) +  u(t\V{e}-\V{q})\} + C(0) \\
                                      & = & \max_{\V{s}}\{u(\V{s}) -s_i\}  + C(0).
\end{eqnarray*}
Then, we take the maximum among all scenarios to conclude the proof. \end{proof}
As a corollary of the above theorem, we have that:
\begin{corollary}
Computing the worst case loss bound for the SCPM is a convex
optimization problem. Furthermore, a necessary and sufficient
condition on utility function $u(\cdot)$ in order to guarantee a
bounded loss is that the difference $u(\V{s})-s_i$ is bounded from
above.
\end{corollary}
Below, we illustrate the application of the above theorem through some examples. Detailed proofs for these examples are available in
Appendix \ref{app:examples}.
\begin{example}
For $u(\V{s}) = -b \log\left(\sum_i \exp(-s_i/b)\right)$, we can compute $B=0$, $C(0)=b\log N$,
giving worst case loss as $b\log N$.
\end{example}
\begin{example}
For $u(\V{s}) = \frac{\V{e}^T\V{s}}{N}-\frac{1}{4b}\V{s}^T(I-\frac{\V{e}\V{e}^T}{N})\V{s}$,
we can derive $B=b(1-\frac{1}{N})$, $C(0)= 0$ giving bound $\frac{(N-1)b}{N}$.
\end{example}
Later in Section \ref{sec:unified}, we demonstrate that the above
two choices of utility functions are equivalent to the LMSR and the
Quadratic market scoring rules respectively. Observe that the
derived bounds match those known in the literature for these scoring
rules \cite{yiling}.

\begin{example}
For the Log-SCPM, $u(\V{s}) = \sum_i \theta_i \log(s_i)$, we can
show that $B$ is unbounded by using $s_1=1, s_i =\alpha$, and
letting $\alpha \rightarrow \infty$.
$$B \ge \lim_{\alpha \rightarrow\infty} \ \   \theta_1 \log 1 + \sum_{i\ne1} \theta_i \log(\alpha) - 1 = \infty$$
and $C(0) = \sum_i \theta_i - \sum_i \theta_i \log \sum_i \theta_i$.
Thus, the worst case loss is unbounded in this case.
\end{example}

\begin{example}
\label{MinUtility}
For $u(\V{s})=\min_i s_i$ , the worst case loss is $0$; since for all values of $(\V{s},t)$: $u(\V{s}) \le s_i$ and $t = u(t\V{e})$.
Observe that here the utility of the surplus profit $\V{s}$ is equal to the minimum or ``worst case" profit. This represents extreme
risk averseness of the market organizer.
\end{example}
The last example provided a glimpse of how the utility function relates to the risk averseness of the market maker.
In Section \ref{risk}, we will further build this intuition for the behavior of the
organizer by recasting our optimization framework as a risk
minimization problem.

\section{Relationship of the SCPM and the MSR}
\label{sec:unified} The Market scoring rules (MSR) form a large
class of popular pari-mutuel mechanisms. In this section, we
demonstrate a strong equivalence between the SCPM and the MSR
markets. Particularly, we show that:
\begin{theorem}
\label{th:MSRequiv}
Any proper market scoring rule with cost function $C(\cdot)$ of (\ref{score-cost}) can be formulated as an SCPM model (\ref{scpm1})
with the {\it `concave'} utility function $u(\V{s}) = -C(-\V{s})$,
and the two models are equivalent in terms of the orders accepted
and the price charged for a submitted order.
\end{theorem}
\begin{theorem}
\label{th:SCPM-proper-score} The SCPM with any utility function
$u(\cdot)$ gives an implicit proper scoring rule, as long as the
utility function has the property that its derivative spans the
simplex $\{\V{r}:\ \V{e}^T\V{r}=1,\ \V{r}\ge 0\}$, that is, for all
vectors $\V{r}$ in the simplex:
\begin{equation}
\label{u-properness-condition}
\nabla u(\V{s}) = \V{r}, \ \ \ \exists\ \V{s}.
\end{equation}
\end{theorem}
Thus, the SCPM framework subsumes the class of proper scoring rule mechanisms. Moreover, a proper scoring rule based market
can be created by simply choosing a utility function $u(\cdot)$ that satisfies condition (\ref{u-properness-condition}).
As we shall demonstrate later in this section, this condition is not difficult to satisfy or validate, thus providing a useful tool to
design market mechanisms that correspond to a proper scoring rule. \\
\newline
We first prove the above theorems for general MSR based market, and
then illustrate with specific examples of the LMSR and the Quadratic
market scoring rules.
\subsection{Equivalent SCPM for MSR (Theorem \ref{th:MSRequiv})}
To establish Theorem \ref{th:MSRequiv}, we use the cost function formulation of market scoring rules discussed in \cite{yiling},
and briefly explained in Section \ref{sec:background-cost-function}.
We first establish the following important properties of the cost
function for {\emph{any}} proper scoring rule:
\begin{lemma}\label{cost_convex}
The cost function $C(\cdot)$ for any proper scoring rule has following properties:
\begin{enumerate}
\item $C(\V{q})$ is a convex function of $\V{q}$
\item For any vector $\V{q}$ and scalar $d$, it holds that: $C(\V{q}+d\V{e})=d+C(\V{q})$
\end{enumerate}
\end{lemma}
\begin{proof}
The proof is referred to Appendix  \ref{app:MSRequiv}. Note that in \cite{yiling}, the second property above was treated as an
assumption based on the principle of no arbitrage. Here, we show that it can actually be derived from the properties of cost
function formulation itself.
\end{proof}

Now, we are ready to prove Theorem \ref{th:MSRequiv}:
\begin{proof}
Using the result in part 1 of Lemma \ref{cost_convex}, clearly the proposed utility function $u(\V{s}) = -C(-\V{s})$ is concave. \\
In light of part 2 of Lemma \ref{cost_convex}, 
we have $C(-\V{s}+z\V{e}) = C(-\V{s}) + z$. Therefore, the proposed utility function $u(\V{s}) = -C(-\V{s}) = z-C(z\V{e}-\V{s})$.
Incorporating $u(\V{s})=z-C(z\V{e}-\V{s})$ in the SCPM model (\ref{scpm1}), we get the following optimization problem:
\begin{equation*}
\begin{array}{lll}\label{variant2}
\mbox{maximize}_{x,z,\V{s}}& \pi x-C(z\V{e}-\V{s})\\
\mbox{subject to } & z\V{e} - \V{a}x - \V{s} = \V{q} \\
&0\leq x\leq l\\
\end{array}
\end{equation*}
Since, $C(\cdot)$ was proven to be convex in part 1 of Lemma \ref{cost_convex}, this is a convex optimization problem with KKT conditions:
\begin{equation*}
 \begin{array}{l}
 \V{p}^T\V{a}+y\geq\pi,\quad x\cdot(\V{p}^T\V{a}+y-\pi)=0,\\
 p_i=\frac{\partial C(\V{q}+\V{a}x)}{\partial q_i},\quad y\cdot(l-x)=0,\quad y \ge 0, 0\leq x\leq l.
 \end{array}
 \end{equation*}
 Thus, $x$ is increased until $x=l$, or the price $\V{p}^T\V{a} = \nabla C(\V{q}+\V{a}x)^T\V{a}$
 becomes greater than the bid price $\pi$. In particular, under continuous charging
 scheme, where a series of optimization problems are solved with small $l=\epsilon$, this
 corresponds to charging the increasing instantaneous price
 $\V{p}(\epsilon)^T\V{a} = \nabla C(\V{q}+\V{a}\epsilon)^T\V{a}$
 until it is greater than the price offered by the bidder.
 For infinitesimally small $\epsilon$,
 $lim_{\epsilon \rightarrow 0} p(\epsilon)_i =  \frac{\partial C}{\partial q_i}$.
 Thus, the orders accepted and price charged become equivalent to the market scoring rule with cost function $C(\V{q})$.
\end{proof}
\subsection{Properness of the SCPM (Theorem \ref{th:SCPM-proper-score})}
In this section, we prove that the SCPM mechanism is proper, that is, it implicitly corresponds to scoring the reported
beliefs with a proper scoring rule, as long as the utility function satisfies the spanning condition (\ref{u-properness-condition}).

\begin{proof}
By definition, a scoring rule $\score(\cdot)$ is proper if and only
if given any outcome distribution $\V{r}$, an optimal strategy of a
selfish trader is to report belief $\V{r}$, that is:
\begin{eqnarray*}
& \V{p^*} \in \arg \max_{\V{p}} \sum_i r_i \score_i(\V{p}) & = \V{r}
\end{eqnarray*}
In cost function based markets like the SCPM, the traders do not
directly report a belief $\V{p}$. Instead, they buy shares $\V{q}$
paying a price that equals to the difference of the cost function,
and thus indirectly reporting the belief as the resulting price
vector $\V{p}$. For these markets, an implicit scoring rule is
defined in the following manner \cite{yiling}:
\begin{equation*}
\begin{array}{ll}
& \score_i(\V{p})=q_i-C(\V{q})+K\mbox{       }\forall i\\
\textrm{where}& p_i=\frac{\partial C}{\partial q_i}\mbox{       }\forall i\\
\end{array}
\end{equation*}
(also refer equation (\ref{score-cost}) in Section
\ref{sec:background-cost-function}). Therefore, the properness
condition in terms of $\V{q}$ is represented as:
\begin{eqnarray}
\label{eq1}
& \V{p^*} := \nabla C(\V{q^*}) = \V{r}& \\
\label{eq2} \textrm {where} & \V{q^*} \in \arg \max_{\V{q}\ge 0} \sum_i
r_i (q_i-C(\V{q}))&
\end{eqnarray}
for all distributions $\V{r}$.

Intuitively, since the traders receive \$1 for each share on the
actual outcome, the profit of traders for outcome state $i$ is
$q_i-C(\V{q})$. Thus, the ``properness'' condition ensures that an
optimal strategy for selfish traders is to buy orders $\V{q}$ so
that the resulting price vector is equal to their actual belief.
Now, the optimality conditions for (\ref{eq2}) are:
$$ r_i - \frac{\partial C(\V{q^*})}{\partial q^*_i} + \eta^*_i =0, \eta^*_i\ge 0, q^*_i \ge 0, \eta^*_iq_i^*=0,\mbox{  }\forall i$$
Thus, condition  (\ref{eq1}) is satisfied if there exists a positive optimal solution to (\ref{eq2}).
As derived in Theorem \ref{th:SCPM-cost-function}, the cost function
of the SCPM mechanism is given by (\ref{costofscpm}).
Therefore, the optimization problem (\ref{eq2}) is equivalent to
$$
\begin{array}{lll}
& \max_{\V{q}\ge 0} & \sum_i r_i (q_i-\min_t \{t-u(t\V{e} - \V{q})\})\\
\Leftrightarrow & \max_{\V{q}\ge 0,t} & \sum_i r_i (q_i- t + u(t\V{e} - \V{q}))\\
\Leftrightarrow & \max_{\V{q}\ge 0,t} & \V{r}^T(\V{q} - t\V{e}) + u(t\V{e} - \V{q})\\
\Leftrightarrow & \max_{\V{s}: \V{s}=t\V{e}-\V{q}, \V{q}\ge 0} &
u(\V{s})-\V{r}^T\V{s}\\
\Leftrightarrow & \max_{\V{s}} & u(\V{s})-\V{r}^T\V{s}
\end{array}
$$
As long as there exists an optimal solution $\V{s}^*$ to the above
problem, we can set $t^*$ as a large positive value and set
$\V{q^*}=t^*\V{e}-\V{s^*} > 0$. Thus, the condition
(\ref{u-properness-condition}), that is, $\nabla u(\V{s})$ spans the
simplex,  ensures the properness.
It is easy to see that this is also a necessary condition.
This proves Theorem \ref{th:SCPM-proper-score}.\end{proof}

A concern however is that the price vector $\V{p^*}$ that maximizes
trader's expected profit may not be unique. This could be either because there are multiple sub-gradients of the cost function
$C(\V{q})$ at optimal $\V{q^*}$ resulting in multiple
price vectors $\V{p^*}$, or because there are multiple optimal
$\V{q^*}$ and they all result in different corresponding price vectors
$\nabla C(\V{q^*})$. This is typically undesirable since in this
case, either buying the orders $\V{q^*}$
associated with the true belief $\V{r}$ is not the only optimal strategy for the traders, or even in the case that the traders acquire $\V{q^*}$, the market maker is still unable to recover the true belief. This situation is avoided by the concept of strictly proper scoring rules.
A scoring rule is called
{\em ``strictly proper''} if the {\em only} optimal strategy for
traders is to honestly report the belief \cite{winkler1969}.
In terms of our market mechanism, it means that the optimal price
vector $\V{p^*}$ that satisfies condition $(\ref{eq1})-(\ref{eq2})$
must be unique. Since $u(\cdot)$ is concave, it is easy to see that a
sufficient condition to ensure strict properness in the SCPM is that $u(\cdot)$ is a smooth
function, that is, $\nabla u(\cdot)$ is continuous (over the simplex).

Next, we illustrate the equivalence between the SCPM and proper
scoring rules using some popular mechanisms as examples.
\subsection{Examples from existing mechanisms}
\begin{example}
The LMSR market maker is equivalent to the SCPM framework with
utility function $u(\V{s}) = -C(-\V{s})=
-b\log{(\sum_i{e^{-s_i/b}})}$. This scoring rule is known to be
strictly proper \cite{Hanson}. Note that our condition for
properness is satisfied as well since $u(\cdot)$ is smooth and
$$\nabla u(\V{s}) =  \left[\frac{e^{-s_i/b}}{\sum_i e^{-s_i/b}}\right]$$
which clearly spans the simplex.
\end{example}
\begin{example}
A market maker using Quadratic-Scoring-Rule is equivalent to the
SCPM framework with utility function $u(\V{s}) =-C(-\V{s})=
\frac{\V{e}^T\V{s}}{N}-\frac{1}{4b}\V{s}^T(I-\frac{\V{e}\V{e}^T}{N})\V{s}$.
This scoring rule is known to be strictly proper \cite{yiling}. Our
condition for properness is satisfied since $u(\cdot)$ is smooth and
$$\nabla u(\V{s}) = \left[ \frac{1}{N} + \frac{\bar{s}-s_i}{2b}\right]$$
where $\bar{s} = \V{e}^T\V{s}/N$. Thus, for any $\V{r}$ in simplex, we can set $s_i=-2br_i$ to get $\nabla u(\V{s})=\V{r}$
\end{example}
\begin{example}
For the Log-SCPM \cite{peters}, $u(\V{s})=\sum_i \theta_i \log s_i$,
$\nabla u(\V{s})_i = \theta_i/s_i$, which clearly spans the interior
of the simplex for any positive $\V{\theta}$. Also, $u(\cdot)$ is
smooth, thus this mechanism is strictly proper.
\end{example}
\begin{example}
For $u(\V{s})=\min_i s_i$, the set of sub-gradients at $\V{s}=\V{e}$
is the convex hull of orthogonal vectors $\{\V{e}_i\}_{i=1,\ldots,
n}$ where $\V{e}_i$ denotes a vector with $1$ at position $i$ and
$0$ elsewhere. This convex hull is exactly the simplex. Thus, the
scoring rule in the SCPM with this utility is proper but not
strictly proper.
\end{example}
\begin{example}
Consider a linear utility function $u(\V{s}) = \V{c}^T\V{s}$. This mechanism is not proper, since the derivative of the function is a constant
vector, and does not span the simplex.
\end{example}
So far, we established the SCPM as a general market mechanism that
includes all scoring rules which are proper and possess common
desired properties. A natural question would be: what is the
difference among the different utility objective functions adopted
in the SCPM? Next, we will show that
they represent different risk measures for the market maker. Thus, when using this
model to accept incoming orders the market maker is actually taking rational decisions with
respect to a specific risk attitude defined in terms of $u(\cdot)$.


\section{Risks for the Market Maker} \label{risk}
Each time he or she is offered an order, the market maker must
consider the risks involved in accepting it. This is due to the fact
that the monetary return generated from the market depends on the
actual state outcome. In the earlier pari-mutuel market introduced
in~\cite{lange05}, this risk was effectively handled in terms of
maximizing the worst case return generated by the market relative to
the set of outcomes (i.e., $u(\V{s})=\min_i s_i$, refer
Example~\ref{MinUtility}). Unfortunately,  this risk attitude is
somewhat limiting as it will create a market which is likely to
accept very few orders and extract little information. In what
follows, we consider the return generated by the market to be a
random variable $Z$ and demonstrate that, when accepting orders
based on the SCPM with a non-decreasing utility function, the market
maker effectively takes rational decisions with respect to a risk
attitude. We use duality theory to gain new insights about how this
attitude relates to the concept of prior belief about the true
probability of outcomes.

\subsection{The SCPM Markets and the Convex Risk Measures}
In a finitely discrete probability space $(\Omega,\mathcal{F})$, the
set of random variables $\mathcal{Z}$ can be described as the set of
functions $Z : \Omega \rightarrow \Re$. A convex risk measure on the set $\mathcal{Z}$
is defined as follows:

\begin{definition}
When the random outcome $Z$ represents a return, a risk measure is a
function $\rho: \mathcal{Z} \rightarrow \Re$ that describes one's
attitude towards risk as : random return $Z$ is preferred to $Z'$ if
$\rho(Z) \leq \rho(Z')$ . Furthermore, a risk measure is called
convex if it satisfies the following:
\begin{itemize}
\item \textit{Convexity} : $\rho(\lambda Z + (1-\lambda) Z') \leq \lambda
\rho(Z) + (1-\lambda)\rho(Z')$, $\forall \;Z,Z'\in \mathcal{Z}$, and
$\;\forall\; \lambda \in [0,1]$
\item \textit{Monotonicity} : If $Z,Z'\in\mathcal{Z}$ and $Z\geq Z'$ then
$\rho(Z) \leq \rho(Z')$
\item \textit{Translation Equivariance} : If $\alpha \in \Re$ and
$Z\in \mathcal{Z}$, then $\rho(Z+\alpha) = \rho(Z)-\alpha$
\end{itemize}
\end{definition}

Convex risk measures are intuitively appealing. First, even in a
context where the decision maker does not know the probability of
occurrence for the different outcomes, it is still possible to
describe a risk function $\rho(Z)$. The three properties of convex
risk measures are also natural ones to expect from such a function.
Convexity states that diversifying the returns leads to lower risks.
Monotonicity states that if the returns are reduced for all outcomes
then the risk is higher. And finally, translation invariance states
that if a fixed income is added to random returns then it is
irrelevant wether this fixed income is received before or after the
random return is realized. We refer the reader
to~\cite{follmer02:cvxRiskMeas} for a deeper study of convex risk
measures.

Next, we formulate the SCPM model for prediction markets as a convex
risk minimization problem. In context of prediction markets, the
random return $Z$ will represent the revenue for the market
organizer, which depends on the actual outcome of the random event
in question. Let $\V{q}$ represent the total orders held by the
traders, and $c$ represent the total money collected so far from the
traders in the market. Since the market organizer has to pay $\$1$
for each accepted order that matches the outcome, his revenue for
outcome state $i$ is $c-q_i$. When a new trader enters with a bid of
$\pi$, based on the number of accepted orders $x$, the total revenue
for state $i$ is given by $(c - q_i + \pi x - a_i x)$. The risk
minimization model seeks to choose the number of accepted orders $x$
to minimize the risk on total revenue. Below, we formally show that
the SCPM model is equivalent to a convex risk minimization model.

\begin{theorem}\label{theoremConvexRisk}
Let $\Omega = \{\omega_1,\omega_2,...,\omega_m\}$, $\V{Z}\in\Re^m$
be the vector representation of $Z$ such that $\V{Z}_i =
Z(\omega_i)$, and $Z^x(\omega_i) = c - q_i + \pi x - a_i x$. Then,
given that $u(\cdot)$ is non-decreasing, the SCPM optimization model
(\ref{scpm1})
is equivalent in terms of set of optimal solutions for $x$ to the
risk minimization model
\begin{eqnarray*}
\minimize_{x} && \rho( Z^x )\\
\subto && 0\leq x\leq l
\end{eqnarray*}
with convex risk measure $\rho(Z) = \min_t \{t -u(\V{Z}+t\V{e})\}$.
\end{theorem}

\begin{proof}
The equivalence can be obtained by first eliminating $\V{s}$ in
(\ref{scpm1}), and then performing a simple change of variable $t =
z - \pi x - c$:
\begin{equation*}
\begin{array}{rcl}
& & \max_z \pi x - z + u\left( z \V{e} - \V{a}x - \V{q}\right) \\
&=& \max_{t} -t + u\left( t\V{e} + (\pi x + c)\V{e}- \V{a}x -\V{q}\right) -c\\
&=& - \min_t \{t - u\left( \V{Z}^x + t \V{e}\right)\}-c \\
&=& -\rho(Z^x)-c \;\; .
\end{array}
\end{equation*}
Since maximizing $-\rho(Z^x)-c$ over $x$ is equivalent to minimizing
$\rho(Z^x)$ in terms of optimal solution set, the equivalence
follows directly.

It remains to show that, when $u(\cdot)$ is concave and
non-decreasing, the proposed measure satisfies the three properties
(convexity, monotonicity, and translational equivariance) of a
convex risk measure. The convexity and the monotonicity follow directly
from concavity and monotonicity of $u(\cdot)$. We refer the reader
to Appendix C for more details on this part of the proof.
\end{proof}

\begin{remark} More importantly, Theorem \ref{theoremConvexRisk} essentially shows that {\em any} convex risk
measure $\rho(Z)$ can potentially be used to create a version of the
SCPM market which accepts orders according the risk attitude described by
$\rho(Z)$. This is achieved by simply choosing the utility function
$u(\V{s}) = -\rho(Y^{\V{s}})$ where $Y^{\V{s}}:\Omega\rightarrow \Re$ is a random
variable defined as $Y^{\V{s}}(\omega_i)=s_i$. Such a constructed
$u(\cdot)$ is necessarily concave and increasing.
\end{remark}

\subsection{Risk attitude through the belief space}
We just showed that the SCPM actually represents a risk minimization
problem for the market maker when $u(\cdot)$ is non-decreasing. In
fact, we can get more insights about the specific risk attitude by
studying the dual representation of risk measure $\rho(Z)$:
\begin{eqnarray}
\rho(Z) = \min_{\V{p}\in\{\V{p}|\V{p}\geq0,\sum_i p_i = 1\}} \Expect_{\V{p}}[Z] +
\Loss(\V{p}) \ \ ,
\end{eqnarray}
where $\Loss(\V{p}) = \max_{\V{s}} u(\V{s}) - \V{p}^T \V{s}$, and
$\Expect_{\V{p}}[Z] = \sum_i p_i \V{Z}_i$. We refer the reader
to~\cite{follmer02:cvxRiskMeas} for more details on the equivalence
of this representation. Note that $\rho(Z)$ is evaluated by
considering the worst distribution $\V{p}$ in terms of trading off
between reducing expected return and reducing the penalty function
$\Loss(\V{p})$.

In terms of the SCPM, this representation equivalence leads to the
conclusion that orders are accepted according to:
\begin{eqnarray}
\max_{0\leq x\leq l} \;\;\left(\min_{\V{p}\in\{\V{p} | \V{p}\geq0,\sum_i p_i =
1\}} \sum_i p_i \V{Z}^x_i + \Loss(\V{p}) \right)\ \ .
\end{eqnarray}
In this form, it becomes clearer how $\Loss(\V{p})$ encodes the intents
of the market maker and relates it to his belief about the true
distribution of outcomes. For instance, we know that the first order
is accepted only if
\[\forall\,\V{p}\in\{\V{p} \ | \V{p}\geq0,\sum_i p_i =
1\},\;\,\; \sum_i p_i\V{Z}^x_i \geq -(\Loss(\V{p})-\Loss(\hat{p}))
\;,\] where $\hat{p}=\argmin_{\V{p}\in\{\V{p}|\V{p}\geq0,\sum_i p_i
= 1\}} \Loss(\V{p})$. For any given $\V{p}$, the penalty
$\Loss(\V{p})-\Loss(\hat{p})$ therefore reflects how much the
market maker is willing to lose in terms of expected returns in the
case that the true distribution of outcomes ends up being $\V{p}$.
It is also the case that after accepting $\V{q}$ orders, the
distribution described by $\V{p^*} =\argmin_{\V{p}} \V{p}^T (c
\V{e}-\V{q}) + \Loss(\V{p})$ is actually the vector of dual prices
computed in the SCPM. In other words, the price vector in the SCPM
market reflects
the distribution that is being considered as the outcome distribution by the market organizer in order to determine his expected return.
This confirms the interpretation of prices as a belief consensus
on outcome distribution generated from the market.

As we will see next, the function $\Loss(\V{p})$ will typically be chosen so that $\Loss(\V{p})-\Loss(\hat{p})$ is large
if $\V{p}$ is far from $\hat{p}$, and $\hat{p}$ will reflect a prior belief of the market organizer. That is, the market organizer is
willing to lose on the expected return in order to learn a distribution $\V{p}$ that is very different from his prior belief.
This is in accordance with the fact that the market is being organized as a prediction market rather than a pure financial market,
and one of the goals of market organizer is to learn beliefs even if at some risk to the generated returns.

We make the above interpretations clearer through the following
examples.
\begin{example}\label{MinSCPMrisk}
The utility function $u(\V{s}) = \min_i \{s_i\}$ corresponds to
cost $\Loss(\V{p}) = \min_i \{s_i\} - \V{p}^T\V{s} = 0$ for all $\V{p}$. That is, the market organizer is purely maximizing
his worst case return.
\end{example}

\begin{example}
For the LMSR, $u(\V{s}) = - b \log \sum_i \exp{(-s_i/b)}$, which is
equivalent to using $\Loss(\V{p})$ as the Kullback-Leibler
divergence of $\V{p}$ from uniform distribution
$\Loss_{\mbox{KL}}(\V{p}\,;\,U)$. This is minimized at $\hat{p}=U$
reflecting a uniform prior.
The corresponding risk measure is also known as the
Entropic Risk Measure and its level of tolerance is measured by $b$.
\end{example}

\begin{example}\label{ExampleSCPMRisk}
The Log-SCPM uses $u(\V{s}) = \sum_i \theta_i \log(s_i)$, which is
equivalent to choosing the penalty function to be the negative
log-likelihood of $\V{p}$ being the true distribution given a set of
observations described by the vector $\V{\theta}$. More specifically,
$\Loss_{\mbox{LL}}(\V{p}\,;\,\V{\theta}) = -\log \left(\prod_i
p_i^{\theta_i}\right)+K$, which is minimized at
$\hat{p}_i=\frac{\theta_i}{\sum_i \theta_i}$ and tolerance to risk is
measured through $\sum_i \theta_i$.
\end{example}

These examples illustrate how the risk minimization representation
provides insights on how to choose $u(\cdot)$. In the case of the
Min-SCPM (Example \ref{MinSCPMrisk}), 
the associated penalty function leads to a market where trades that
might generate a loss for the market maker are necessarily rejected.
Hence, the traders have no incentive for sharing their belief. On
the other hand, both the LMSR and the Log-SCPM are mechanism that
will accept orders leading to negative expected returns under a
distribution $\V{p}$, as long as this distribution is \quoteIt{far
enough} from $\hat{p}$. Effectively, a trader with a belief that
differs from $\hat{p}$ will have his order accepted given that he
submits it early enough. In practice, choosing between the LMSR and
the Log-SCPM involves determining whether the Kullback$-$Leibler
divergence or a likelihood measure better characterizes the market
maker's commitment to learning the true distribution.


\section{Discussion} \label{discussion}

In this work, we introduced a unified convex optimization framework
for constructing prediction market mechanisms. We first showed that
in this new framework, the pricing mechanism always allows truthful
bidding (in a myopic sense) to be an attitude that is optimal for
the traders. Also, the pricing mechanism always leads to prices that
can be computed efficiently using a convenient convex cost function
formulation. We showed how the original SCPM mechanism and
mechanisms that are based on scoring rules could be cast in this
unifying framework. These mechanisms actually differ only in terms of the
choice of utility function. This fact led to the analysis of
properties of this utility function which are of particular interest
when designing a prediction market. We first proposed a way to
compute the potential worst case loss obtained from the market in
this framework. We also discussed conditions on the utility function
to ensure `properness' of the mechanism inline with the definition
of a proper scoring rule. Later, we showed that when selecting a
non-decreasing utility function, the market maker is implicitly
defining his risk attitude with respect to the potential revenues
obtained from the market.\\
\newline
\newcommand{\tabletextone}[1]{{\small #1}}
\newcommand{\tabletexttwo}[2]{$\begin{array}{c}
\textrm{\small #1}\\
\textrm{\small #2}
\end{array}$}
\newcommand{\tabletext}[1]{{\small #1}}
\begin{center}
\begin{table*}[t]
\label{comparison-table}
\caption{A summary of properties of various market mechanisms}
\begin{tabular}{|l|c|c|c|c|c|c|}
\hline
&$u(\V{s})$&\tabletextone{Truthful}{}&\tabletexttwo{Worst}{Cost}&
\tabletexttwo{Convex}{Risk Measure}
&$L(\V{p})$& \tabletextone{Properness}\\
\hline
\tabletextone{LMSR}&$-b\log(\sum_i\exp(-\frac{s_i}{b}))$&Yes& $b\log{N}$&Yes&$b\Loss_{\mbox{KL}}(\V{p}||U)$&\tabletext {Strictly Proper}\\
\tabletextone{Log-SCPM}&$b \sum_i\log(s_i)$&Yes&$\infty$&Yes&$b\Loss_{\mbox{LL}}(\V{p}||U)$& \tabletext{Strictly Proper}\\
\tabletextone{Min-SCPM}&$\min_i s_i$& Yes&$0$&Yes&0&\tabletext{Proper}\\
\tabletextone{Quad. Scoring Rule}&$\frac{1}{N}\V{e}^T\V{s} - \frac{1}{4b}\V{s}^T(I-\frac{1}{N}\V{e}\V{e}^T)\V{s}$&Yes&$b\frac{N-1}{N}$&No&-&\tabletext{Strictly Proper}\\
\oldExpSCPM{\tabletextone{Exponetial-SCPM}&$b(1-\frac{1}{N}\sum_i e^{-s_i})$& Yes&$\log{N}$&Yes&$\sum_i p_i \log p_i$& \tabletext{Strictly Proper}\\}
\newExpSCPM{\tabletextone{Exponetial-SCPM}&$b(1-\frac{1}{N}\sum_i \exp(-\frac{s_i}{b}))$& Yes&$b\log{N}$&Yes&$b\Loss_{\mbox{KL}}(\V{p}||U)$& \tabletext{Strictly Proper}\\}
\removeLinSCPM{\tabletextone{Linear-SCPM}&$\left\{\begin{array}{cl}\frac{1}{N}\sum_i \min(0,s_i)& \mbox{if $s\geq-be$}\\ -\infty&\mbox{o.w.}\end{array}\right.$& Yes&$b\frac{N-1}{N}$&Yes&$\frac{b}{2}\|\V{p}-U\|_1$& \tabletext{Proper}\\}
\tabletextone{Quad-SCPM}&$\max_{\V{v}\leq\V{s}} \frac{1}{N}\V{e}^T\V{v} - \frac{1}{4b}\V{v}^T\V{v}$&Yes&$b\frac{N-1}{N}$&Yes&$b\|\V{p}-U\|_2^2$ & \tabletext{Strictly Proper}\\
\hline
\end{tabular}
\end{table*}
\end{center}
We believe these properties to be very valuable for selecting the
most effective market mechanism in a given application. Table 1
summarizes the conclusions we derived for a set of popular
mechanisms: the LMSR, the original SCPM (Log-SCPM), the riskless
Market (Min-SCPM), and the market based on the quadratic scoring
rule. The table relates truthfulness and properness of the
mechanisms to the worst case loss and the relationship to market
maker's risk attitude. We believe that these results can provide
guidance in designing cost effective prediction markets. The table
also introduces new versions of the SCPM model described in
Example~\ref{expSCPM}\removeLinSCPM{, Example~\ref{linearSCPM},}
and~\ref{quadSCPM}. These models are valuable since they extend very
naturally the landscape of structures for characterizing the market
maker's risk attitude through the definition of penalty functions
$\Loss(\V{p})$.
\newExpSCPM{
\begin{example}
\label{expSCPM} Consider the \quoteIt{Exponential-SCPM} obtained from
using the following utility:
\[u(\V{s}) = b(1-\frac{1}{N}\sum_i e^{-s_i/b})\]
This utility function is concave, non-decreasing and separable.
It has a bounded worst case loss equal to $b\log N$. And, the
function $\Loss(\V{p})$ is the Kullback-Leibler divergence of $\V{p}$ from
uniform distribution $\Loss_{\mbox{KL}}(\V{p}\,;\,U)$.
Using the pricing scheme
described in Section~\ref{sec:truthfulness}, it leads to a
prediction market that is myopically truthful. Also, the
corresponding scoring rule is strictly proper. And, the orders can
be priced using convex cost function $C(\V{q}) = \log \sum_i \frac{e^{q_i/b}}{N}$.
\end{example}
The above example shows that we can achieve properties same as the
LMSR using an alternative  {\it separable} utility function in the
SCPM. }
Our unified analysis of the SCPM model also allows us to suggest a
modification to the prediction market that uses a quadratic scoring
rule. Although this rule is known to be myopically truthful, in
practice a market maker that uses this rule needs to explicitly
restrict prices to be between [0,1] at all times. A solution to this
problem is to use an SCPM market with the following semi-quadratic non-decreasing
utility function:
\begin{example}
\label{quadSCPM}
Consider the \quoteIt{Quad-SCPM} obtained from using the following
utility:
\[u(\V{s})  = \max_{\V{v}\leq \V{s}} \frac{1}{N}\V{\theta}^T\V{v} - \frac{1}{4b}\V{v}^T\V{v}\]
for some $\V{\theta}$ such that $\sum_i\theta_i = 1$.
This utility function is non-decreasing and concave which ensures
that resulting prices are non-negative and sum to $1$. It has
bounded worst case lost given by $b(\|\V{\theta}\|_2^2 +
1 - 2\min_i \theta_i$). And, the distance from the prior
$\hat{p} = \V{\theta}$ is measured by $\Loss(\V{p}) = b
\|\V{p}-\V{\theta}\|_2^2$, that is the 2-norm distance. The resulting
prediction market is myopically truthful and leads to orders that
can be priced using the cost function:
\[C(\V{q}) = \min_{t,\V{v}\leq t\V{e}-\V{q}} t - \V{\theta}^T\V{v} + \frac{1}{4b}\V{v}^T\V{v}\;.\]
which requires solving a quadratic program.
Also, the corresponding scoring rule is strictly proper.
\end{example}
The above market is closely related to the market associated with
the quadratic scoring rule, since $C(\V{q})$ reduces to the popular
quadratic cost function when the constraint $\V{v}\leq t\V{e}
-\V{q}$ is replaced with $\V{v}=t\V{e}-\V{q}$. However, the slight
modification ensures that prices are non-negative, and has an
intuitive interpretation in terms of distance to the prior belief.

Detailed proofs for the properties summarized in Table 1 are available in Appendix \ref{app:examples}.



\appendix

\section{Properties of pricing scheme}
\subsection{Proof of Lemma \ref{price-properties}}
\label{app:price-properties}

Consider the KKT condition for (\ref{scpm2}).
\begin{equation}
\begin{array}{ll}\label{KKT2}
\V{e}^T\V{p}&=1\\
p_i&=\nabla u(\V{s})_i\\
\V{p}^T\V{a}+y&\geq\pi\\
x\cdot(\V{p}^T\V{a}+y-\pi)&=0\\
y&\geq 0\\
y\cdot(\epsilon-x)&=0\\
\V{a}x+ \V{q} +\V{s}&=z\V{e}\\
\end{array}
\end{equation}

{\it 1., 2. } The first and second assertions follow from the top two equations in
(\ref{KKT2}), and the assumption that $u(\cdot)$ is non-decreasing.\\
{\it 3. } The consistency of price follows from constraints 3-6 in (\ref{KKT2}).\\
{\it 4. } Let $(x_1,z_1,\V{s_1}),(x_2,z_2,\V{s_2})$ be the optimal solution       
of (\ref{scpm2}) for the case of $\epsilon=\epsilon_1$ and
$\epsilon_2$ respectively, where $\bar{x}>\epsilon_2>\epsilon_1$.
Observe that from the problem formulation, $x_2>x_1$. Since $u$ is
concave, we know that
\begin{equation*}
\begin{array}{l}
u(z_2\V{e}-\V{q}-x_2\V{a})-u(z_1\V{e}-\V{q}-x_1\V{a})\\
\ \ \ \ \leq \nabla u(z_1\V{e}-\V{q}-x_1\V{a})^T((z_2-z_1)e-(x_2-x_1)\V{a})\\
u(z_1\V{e}-\V{q}-x_1\V{a})-u(z_2\V{e}-\V{q}-x_2\V{a}) \\
\ \ \ \ \leq \nabla u(z_2\V{e}-\V{q}-x_2\V{a})^T((z_1-z_2)e-(x_1-x_2)\V{a})\\
 \end{array}
 \end{equation*}
Therefore by adding the above inequalities, we get
\begin{equation*}
(\nabla u(z_1\V{e}-\V{q}-x_1\V{a})-\nabla
u(z_2\V{e}-\V{q}-x_2\V{a}))^T((z_2-z_1)e-(x_2-x_1)\V{a})\geq 0\\
\end{equation*}                                                                 
However, by the KKT conditions (refer (\ref{KKT2})), we have
\begin{equation*}
\begin{array}{l}
\nabla u(z_1\V{e}-\V{q}-x_1\V{a})^T\V{e}=1, \nabla u(z_2\V{e}-\V{q}-x_2\V{a})^T\V{e}=1\\
\end{array}
\end{equation*}
Also, $x_2>x_1$. Therefore, we have
$$(\nabla u(z_2\V{e}-\V{q}-x_2\V{a})-\nabla
u(z_1\V{e}-\V{q}-x_1\V{a}))^T\V{a}\geq 0$$ which means
$\V{p}^T(\epsilon_2)\V{a}\geq
\V{p}^T(\epsilon_1)\V{a}$. This proves the claim.\\
{\it 5. } To prove this claim, we use the following result from real
analysis: {\it a bounded increasing or decreasing function on a
finite interval is integrable}. Using the observation made in part
4, the proof follows.

\begin{remark} For ease of presentation, in above proof we assumed $u$ to be differentiable. Otherwise, the proof
still holds with the gradients replaced by sub-gradients.
\end{remark}
\section{Equivalence of the MSR and the SCPM}
\label{app:MSRequiv}
\subsection{Proof of Lemma \ref{cost_convex}}
\begin{enumerate}
\item {\it $C(\V{q})$ is a convex function of $\V{q}$}
\begin{proof}
To prove $C(\V{q})$ is convex, it suffices to
show that for any $\V{q}_0$ and $\V{q}_1$, $C(\V{q}_0+\lambda \V{q}_1)$ is convex in
$\lambda$. We have the following:
\begin{equation*}
\begin{array}{ll}
\frac{dC}{d\lambda}&=\V{q}_1\cdot\nabla
C|_{(\V{q}_0+\lambda \V{q}_1)}\\
&=\V{q}_1\cdot p(\V{q}_0+\lambda \V{q}_1)\\
\end{array}
\end{equation*}
Therefore, it suffices to show that $\V{q}_1\cdot \V{p}(\V{q}_0+\lambda \V{q}_1)$ is
increasing in $\lambda$.\\
\newline
We will denote $\V{p}(\V{q}_0+\lambda \V{q}_1)$ by $\V{p}(\lambda)$ and
$C(\V{q}_0+\lambda \V{q}_1)$ by $C(\lambda)$. By the first condition of
(\ref{score-cost}) and the properness of $\score$, we have for any
$\lambda_2>\lambda_1$
\begin{equation*}
\begin{array}{l}
\sum_i
p_i(\lambda_1)(\V{q}_0+\lambda_1\V{q}_1)_i-C(\lambda_1) \\
\hspace{1in} = \ \ \sum_ip_i(\lambda_1)\score_i(p(\lambda_1))-K\\
\hspace{1in} \geq \ \ \sum_ip_i(\lambda_2)\score_i(p(\lambda_1))-K\\
\hspace{1in} \geq \ \ \sum_i p_i(\lambda_2)(\V{q}_0+\lambda_1\V{q}_1)_i-C(\lambda_1)\\
\end{array}
\end{equation*}
So, we are left with the following relation:
\begin{equation}
\begin{array}{ll}\label{1}
\sum_i p_i(\lambda_1)(\V{q}_0+\lambda_1\V{q}_1)_i&\geq\sum_i
p_i(\lambda_2)(\V{q}_0+\lambda_1\V{q}_1)_i\\
\end{array}
\end{equation}
 Similarly,
\begin{equation}
\begin{array}{ll}\label{2}
\sum_i p_i(\lambda_2)(\V{q}_0+\lambda_2\V{q}_1)_i\geq\sum_i p_i(\lambda_1)(\V{q}_0+\lambda_2\V{q}_1)_i\\
\end{array}
\end{equation}
Then (\ref{1})-(\ref{2}) yields
\begin{equation*}
\begin{array}{l}
\sum_i(\lambda_2-\lambda_1)(p_i(\lambda_2)-p_i(\lambda_1))q_{1_i}\geq
0\\
\end{array}
\end{equation*}
\newline
Thus, $C(\V{q})$ is convex in $\V{q}$ for every cost function corresponding
to a proper scoring rule.
\end{proof}
\item
{\it For any vector $\V{q}$ and scalar $d$:
\begin{equation*}
\begin{array}{l}\label{property}
C(\V{q}+d\V{e})=d+C(\V{q})\mbox{      }\forall d
\end{array}
\end{equation*}
}
\begin{proof}
We prove by contradiction. If there exists $\V{q}$ and $d$ such that
$C(\V{q}+d\V{e})>d+C(\V{q})$.
Then we set $\bar{q}=\V{q}+d\V{e}$. Then
$\score_i(\V{p})-\score_i(\bar{p})=(q_i-\bar{q}_i)-(C(\V{q})-C(\bar{q}))>0$ for all
$i$, which is impossible since there must be at least one $p'_i>p_i$
and $\score_i$ is increasing function of $p_i$. Similarly, we can prove a
contradiction for $C(\V{q}+d\V{e}) < d+C(\V{q})$.
\newline
\end{proof}
 Therefore, we have now proven both parts of Theorem
\ref{cost_convex}.
\end{enumerate}

\section{Convexity of risk measure $\rho(\V{Z})$}
\begin{itemize}
\item {\em Convexity}:
Since $\rho(Z)=\min_{t}\{t-u(t\V{e}+\V{Z})\}$, and $u(\cdot)$ is concave. We know
that $\rho(Z)$ is convex.
\item {\em Monotonicity}:
The monotonicity also simply results from the monotonicity of
$u(\cdot)$. Given that $\V{Z} \succeq \V{Z'}$ then
\begin{eqnarray*}
\rho(Z) \;=\; \min_t t -u(\V{Z}+t\V{e}) \;\leq\; \min_t t -u(\V{Z'}+t
\V{e}) \;\\
=\; \rho(Z') \;\;,
\end{eqnarray*}
since the inequality is true for any fixed value of $t$.

\item{\em Translation equivariance}:
Finally, translation equivariance can be simply demonstrated with a
change of variable $t'=t+\alpha$ :
\begin{equation*}
\begin{array}{ll}
\rho(Z+\alpha) \;&=\; \min_t t
-u(\V{Z}+(\alpha+t) \V{e}) \;\\
&=\; \min_{t'} t'-\alpha - u(\V{Z} + t')
\;\\
&=\rho(Z) - \alpha\\
\end{array}
\end{equation*}
\end{itemize}

\section{Properties of various SCPM mechanisms}
\label{app:examples}
\subsection{Properties of the LMSR}

\begin{itemize}
\item{\em Worst Case Loss}: First, one can show that $C(0)=b\log{N}$
from: \begin{equation*}
\begin{array}{ll}
C(0)&=\min_t t+b\log(\sum_i(\exp(-\frac{t}{b})))\\
&=b\log{N}
\end{array}
\end{equation*}
 Then, we can verify that $B =
0$.
\begin{equation*}
\begin{array}{ll}
B& = \max_i\;\max_{\V{s}}\; -b\log\left(\sum_j\exp(-\frac{s_j}{b})\right)
-
s_i\\
 &= \max_i\; -b\log(\exp(-\frac{s^*_i}{b}))-s^*_i\;\\
 & =\;0\;.
\end{array}
\end{equation*}
\item{\em Risk Attitude}: Below is the derivation for a more general function $u(\V{s}) = -
b\log( \sum_i \theta_i \exp(-s_i/b))$:
\begin{equation*}
\begin{array}{ll}
&\Loss(\V{p})\\
=& \max_{s} \left(-b\log\left(\sum_i \theta_i \exp(\frac{-s_i}{b})\right) - \sum_i p_i s_i \right)\\
=& b\max_{\V{v}} \left(-\log\left(\sum_i \exp(v_i) \right)-\sum_i p_i (\log(\theta_i) - v_i) \right)\\
=& b \left(  \max_{\V{v}} \sum_i p_i v_i- \log\left(\sum_i \exp(v_i) \right)\right)  -b \sum_i p_i \log(\theta_i)\\
=& b \left( \sum_i p_i \log(p_i) - \sum_i p_i \log(\hat{p_i})\right) - b\log(\alpha)\\
=&b \Loss_{\mbox{KL}}(\V{p}\,||\,\hat{p}) - b\log(\alpha) \ \ ,
\end{array}
\end{equation*}
where we first used the replacement $v_i = - s_i/b +
\log(\theta_i)$, then solved analytically the maximization in terms
of $\V{v}$ (see~\cite{boyd:cvxOpt}, Example 3.25 for details), then
again replaced $\alpha = \sum_i \theta_i$ and $\hat{p} =
\frac{1}{\alpha} \V{\theta}$.
Thus, $\Loss(\V{p})$ is measured in terms of Kullback-Leibler divergence
$\Loss_{\mbox{KL}}(\V{p}\,||\,\frac{\theta}{\alpha})$.
$\V{p}=\frac{\theta}{\alpha}$ minimizes
$\Loss(\V{p})$, while the tolerance to risk is measured by $b$.
 In the
market that was popularized by Hanson, $\V{\theta} = \V{e}$, thus the market
implicitly assumes  a uniform prior.
\end{itemize}

\subsection{Properties of the Log-SCPM}

\begin{itemize}
\item{\em Worst Case Loss}: The worst case loss is unbounded since
$C(0)=(\sum_i \theta_i)(1-\log\sum_i\theta_i)$ and $B = \infty$.
Specifically,
$$C(0)\;=\;\min_t t-\sum_i\theta_i\log(t) = (\sum_i \theta_i)(1-\log\sum_i\theta_i)$$
and
\begin{equation*}
\begin{array}{ll}
B&=\max_i\max_{\V{s}}\sum_j\theta_j\log(s_j) - s_i \;\\
&\geq\lim_{\alpha\rightarrow\infty} \theta_1\log(1)+
\sum_{j\neq 1}\theta_j\log(\alpha) - 1 \\
&=\infty\;,
\end{array}
\end{equation*}
where we restricted the optimization to $s_1=1$ and $s_j=\alpha, \forall j\ne 1$ and
assumed without loss of generality that $\theta_j\neq0$ for some
$j\neq 1$.
\item{\em Risk Attitude} :
The penalty function $\Loss(\V{p})$ can be derived by simple algebra.
Given that $u(\cdot)$ has the form $u(\V{s}) = \sum_i \theta_i
\log(s_i)$, we can show that
\begin{equation*}
\begin{array}{ll}
\Loss(\V{p})&= \max_{\V{s}} (\sum_i \theta_i \log(s_i) - p_i s_i) \;\\
&=\; \sum_i
\left(\theta_i \log(\theta_i/p_i) - \theta_i\right)\\
&= -\sum_i \theta_i \log(p_i) + \sum_i \left(\theta_i \log\theta_i -
\theta_i\right)\\
&=\; \Loss_{\mbox{LL}}(\V{p}\,;\,\V{\theta})) + K\\

\end{array}
\end{equation*}
where $\Loss_{\mbox{LL}}(\V{p}\,;\,\V{\theta}))=-\log(\prod_i p_i^{\theta_i})$. As $\sum_i \theta_i$ increases the curvature of the
negative log-likelihood function increases. Thus, a higher $\sum_i
\theta_i$ leads to more risk tolerance.
\end{itemize}

\subsection{Properties of the Min-SCPM}
\begin{itemize}
\item{\em Worst Case Loss} : Necessarily, $C(0)=\min_t t-\min_i t_i=0$,
while $B\;=\;\max_i\;\max_{\V{s}}\; \{\min_j s_j - s_i\}=0$. Therefore,
the worst case loss is $0$.
\item{\em Risk Attitude} : The penalty function can be derived as
follows:
\begin{equation*}
\begin{array}{ll}
\Loss(\V{p}) &= \max_{\V{s}} \{\min_i s_i - \V{p}^T\V{s}\}\\
&=\max_{\V{s},v\leq s_i\forall i} v - \V{p}^T\V{s}\\
&=\max_{\V{s},v} \min_\lambda v - \V{p}^T\V{s} + \lambda^T(\V{s}-v \V{e})\\
&=\min_\lambda \max_{\V{s},v} v - \V{p}^T\V{s} + \lambda^T(\V{s}-v\V{e})=0\\
\end{array}
\end{equation*}
\end{itemize}

\subsection{Properties of the Quadratic Scoring Rule}
\begin{itemize}
\item{\em Worst Case Loss} : When $u(\V{s}) = \frac{\V{e}^T\V{s}}{N}-\frac{1}{4b}\V{s}^TP\V{s}$ with $P=(I-\frac{1}{N}\V{e} \V{e}^T)$,
we first show that $C(0)=0$ and then that $B = b(1-\frac{1}{N})$ :
\begin{equation*}
\begin{array}{ll}
C(0)&=\min_t t-u(t\V{e})\\
& = \min_t \ t + \frac{1}{4b} t^2 \V{e}^TP\V{e} -t =0\\
\end{array}
\end{equation*}
and
\begin{equation*}
\begin{array}{ll}
B&=\max_{i}\max_{\V{s}}(u(\V{s})-s_i) \\
&= u([-2b,0,\ldots,0]) + 2b \\
&= \frac{-2b}{N} -b(1-\frac{1}{N})+2b\\
& = b(1-\frac{1}{N})\\
\end{array}
\end{equation*}
\item{\em Risk Attitude} : Since the derivative of $u(\V{s})\;=\;\frac{1}{N}\V{e}^T \V{s} -
\frac{1}{4b}\V{s}^TP \V{s}$ is not non-decreasing, this version of the SCPM
does not have an equivalent representation in terms of convex risk
minimization. Specifically, when $\V{s}=(2b,0,0,...0)$
\[\nabla u_1(\V{s}) =-1+\frac{2}{N}< 0\]
\end{itemize}

\newExpSCPM{
\subsection{Properties of the Exponential-SCPM}
\begin{itemize}
\item {\em Cost function}:
\begin{equation*}
\begin{array}{rcl}
C(\V{q}) & = & \min_t t- b(1-\frac{1}{N}\sum_i e^{(-t+q_i)/b})
 \end{array}
 \end{equation*}
Above is minimized at
\begin{equation*}
\begin{array}{l}
1- \frac{1}{N}\sum_i e^{(-t+q_i)/b}=0
\end{array}
 \end{equation*}
Therefore,
\begin{equation*}
\begin{array}{rcl}
C(\V{q}) & = & \log(\frac{\sum_i e^{q_i/b}}{N})\\
 \end{array}
 \end{equation*}
\item{\em Worst Case Loss}:
\begin{equation*}
\begin{array}{rcl}
C(0) & = & 0 \\
B & = & \max_i \max_{\V{s}} b(1-\frac{1}{N}\sum_i e^{-s_i/b})-s_i\\
& = & b\log{N}\\
 \end{array}
 \end{equation*}
Therefore worst case loss is $b\log N$.

\item{\em Risk Attitude}: We simply resolve the definition of
$\Loss(\V{p})$.
\begin{equation*}
\begin{array}{rcl}
\Loss(\V{p})& = & \max_{\V{s}} u(\V{s})-\V{p}^T\V{s}\\
& = & \max_{\V{s}} b(1-\frac{1}{N}\sum_i e^{-s_i/b}) - \V{p}^T\V{s}\\
 & = & b\sum_i p_i\log{(Np_i)}\\
 & = & b\Loss_{\mbox{KL}}(\V{p}||U)\\
\end{array}
 \end{equation*}

\item{\em Properness}:
The function $u(\cdot)$ is smooth, and the gradient $\nabla
u(\V{s})_i = \frac{1}{N}e^{-s_i/b}$. Clearly, the gradient spans the
simplex. Thus, the mechanism is strictly proper.
 \end{itemize}
}

\subsection{Properties of the Quad-SCPM}
\begin{itemize}
\item{\em Worst Case Loss}: First one can show that $C(0)=0$:
\begin{equation*}
\begin{array}{ll}
C(0) &= \min_t \left\{ t- \max_{\V{w}\leq t\V{e}} \{\V{\theta}^T \V{w} - \frac{1}{4b}\V{w}^T\V{w}\}\right\}\\
&=\min_{t,\V{w}} \max_{\V{p}\geq0} t-\V{\theta}^T \V{w} + \frac{1}{4b}
\V{w}^T \V{w} - \V{p}^T(t\V{e}-\V{w})\\
&= \max_{\V{p}\geq0} \min_{t,\V{w}}  t-\V{\theta}^T \V{w} + \frac{1}{4b} \V{w}^T \V{w}
- \V{p}^T(t\V{e}-\V{w})\\
&=\max_{\V{p}\geq0\,\&\, \V{e}^T \V{p} = 1}
-b\|\V{\theta}-\V{p}\|_2^2=0\\
\end{array}
\end{equation*}
 To compute $B$, we have
\begin{equation*}
\begin{array}{ll}
&\max_{i}\max_{\V{s}} u(\V{s})-s_i \\
=& \max_i \;\max_{\V{s},\V{w}\leq \V{s}}\; \V{\theta}^T \V{w} - \frac{1}{4b} \V{w}^T \V{w}
-s_i\\
=&\max_i \;\max_{\V{s},\V{w}} \;\min_{\V{p}\geq 0}\; \V{\theta}^T \V{w} -
\frac{1}{4b} \V{w}^T \V{w} -
s_i + \V{p}^T(\V{s}-\V{w})\\
=&\max_i \;\min_{\V{p}\geq 0} \;\max_{\V{s},\V{w}}\; \V{\theta}^T \V{w} -
\frac{1}{4b} \V{w}^T \V{w} -
s_i + \V{p}^T(\V{s}-\V{w})\\
=&\max_i\; b \left((\theta_i-1)^2+\|\V{\theta}\|_2^2
-\theta_i^2\right)\\
=&\max_i\; b \left(-2\theta_i+1+\|\V{\theta}\|_2^2\right)\\
=& 2b
\left(\frac{1}{2}+\frac{1}{2}\|\V{\theta}\|_2^2 -\min_i\theta_i\right)\\
\end{array}
\end{equation*}
\item{\em Risk Attitude}: We simply resolve the definition of
$\Loss(\V{p})$.
\begin{equation*}
\begin{array}{ll}
 \Loss(\V{p}) &= \max_{\V{s},\V{w}\leq \V{s}}\; \V{\theta}^T \V{w}
-\frac{1}{4b}\V{w}^T \V{w}
-\V{p}^T \V{s}\\
&= \max_{\V{s},\V{w}} \min_{\lambda\geq0}\; \V{\theta}^T \V{w} -\frac{1}{4b}\V{w}^T
\V{w} -\V{p}^T \V{s} + \lambda^T(\V{s}-\V{w})\\
&=b \|\V{p}-\V{\theta}\|_2^2\\
\end{array}
\end{equation*}

\item {\em Properness}:
The utility function $u(\V{s})$ at a given $\V{s} \ge -b\V{e}$ is given by:
\begin{equation*}
\begin{array}{ll}
\max_v & \frac{1}{N}\V{\theta}^T \V{v} - \frac{1}{4b}\V{v}^T \V{v}\\
\textrm{s.t.} & \V{v}\leq \V{s}
\end{array}
\end{equation*}
The partial derivative of this function with respect to $s_i$ is $0$ at $s_i \ge \frac{2\theta_ib}{N}$, and $\frac{\theta_i}{N}-\frac{s_i}{2b}$ at $s_i < \frac{2\theta_ib}{N}$. 
Therefore, gradient $\nabla u(\V{s})_i = \max \{0, \frac{\theta_i}{N}-\frac{s_i}{2b}\}$ which spans the simplex and is continuous on the simplex. Thus, this mechanism is strictly proper.
\end{itemize}

 \end{document}